\documentclass[showpacs,aps]{revtex4}

\usepackage{amssymb}
\usepackage{amsmath}
\usepackage{color}

\def\boldit#1{\mbox{\boldmath$#1$}}

\makeatletter
 \def\eqalign#1{\null\vcenter{\def\\{\cr}\openup\jot\m@th
  \ialign{\strut$\displaystyle{##}$\hfil&$\displaystyle{{}##}$\hfil
      \crcr#1\crcr}}\,}
\makeatother

\begin{document}

\def\n{\nabla}

\def\bn{\boldit{\nabla}}

\def\t{ \times }

\def\c{ \cdot }

\def\d{ \text{div} }

\def\cu{ \text{curl} }

\def\rf{ Eq.~(\ref) }

\def\ct{ \cite }

\def\p{ \partial }

\def\pt{ \partial_{t} }

\def\pa{ \partial_{a} }

\def\eq{\equiv}

\def\a{^{\ast}}

\def\q{\quad}

\def\equ{Eq.~\eqref}

\title{Relativistic least action principle for discontinuous hydrodynamic flows,
Hamiltonian variables, helicity and Ertel invariant}

\author{A.V. KATS \\
Usikov Institute for Radiophysics and Electronics \\ National
Academy of Sciences of Ukraine, \\
  61085, 12 Ak. Proskury St., Kharkiv, Ukraine  \\  e-mail: avkats@online.kharkiv.com}

\begin{abstract}

A rigorous method for introducing the variational principle
describing relativistic ideal hydrodynamic  flows with all
possible types of breaks (including shocks) is presented in the
framework of an exact Clebsch type representation of the
four-velocity field as a bilinear combination of the scalar
fields. The boundary conditions for these fields on the breaks are
found. We also discuss the local invariants caused by the
symmetries of the problem, including relabeling symmetry. In
particular, the generalization of the well-known nonrelativistic
Ertel invariant is presented.
\end{abstract}


\maketitle

%
%
%


\paragraph{Introduction.}
In the paper we discuss some problems related to the ideal
relativistic hydrodynamic (RHD) flows in the framework of the
special relativity. They are pertinent to description of the flows
with breaks in terms of the canonical (Hamiltonian) variables
based upon the corresponding variational principle, and
introducing the generalization of the Ertel invariant. These
subjects are of interest both from the general point of view and
are very useful in solving  the nonlinear problems, specifically,
for the nonlinear stability investigation, description of the
turbulent flows, etc. The necessity to consider the relativistic
flows is motivated by a wide area of applications to the
cosmological problems.



Variational principles for the ideal relativistic hydrodynamic
(RHD) flows are widely discussed in the literature, see, for
instance,  \cite{Schutz70,Brown93,ZakharovKuznetsov97} and
citations therein. As for the nonrelativistic flows, the least
action principle is convenient to formulate in terms of the
subsidiary fields and corresponding velocity representation known
as the Clebsch representation, see
\cite{lamb,GP93,Berd_83,Serrin59,Lin_63,Zakharov71,Salmon82}.
These subsidiary fields can be introduced explicitly by means of
the Weber transformation, \cite{Weber1868}, see also
\cite{lamb,ZakharovKuznetsov97}. Alternatively, they naturally
arise from the least action principle as Lagrange multipliers for
necessary constraints. Using these variables allows one to
describe the dynamics in terms of canonical (Hamiltonian)
variables. The nontrivial character of the Hamiltonian approach is
due to the fact that the fluid dynamics corresponds to the
degenerated case,
see \cite{Dirac64,Gitman86}. 

In the papers \cite{KK_97,Kats_01} it was shown that the
hydrodynamic flows with breaks (including shocks) can be described
in terms of such least action principle, which includes (as
natural boundary conditions) the boundary conditions for the
subsidiary fields. In the nonrelativistic case the triplet of the
subsidiary fields corresponds to the Lagrange labels of the fluid
particles, say, $\mu^{B}$, which are advected by the fluid,
\begin{equation}\label{28_11_03}
d_{t} \mu^{B} = 0, \q d_{t} \eq \pt + \mathbf{v} \c \bn  , \q B =
1, 2, 3,
\end{equation}
where $\mathbf{v}$ denotes three-velocity. These equations along
with the entropy advection and the fluid mass conservation are
assumed as constraints. Corresponding Lagrange multipliers,
$\lambda_{B}$, $\theta$ and $\varphi$, along with $\mu^{B}$ enter
the Clebsch type velocity representation,
\begin{equation}\label{28_11_03_1}
\rho \mathbf{v} = - \rho \bn \varphi - \lambda_{B} \bn \mu^{B} -
\theta \bn s ,
\end{equation}
where $\rho$ and $s$ denote fluid density and the entropy per unit
mass.

\paragraph{Variational principle.}
The relativistic least action principle can be formulated in a
close analogy to the nonrelativistic one. Namely, introduce action
$A$,
\begin{equation}\label{26_04_3_0}
A = \int d^{4} x \, {\cal L} ,
\end{equation}
with the Lagrangian density
    \begin{equation}\label{23_02_04_4}
    \eqalign{
\mathcal{L} =  - \epsilon(n, S) + G J^{\alpha} Q_{, \alpha} \, ,
\\
G = (1, \nu_{B}, \Theta) , \q Q = (\varphi, \mu^{B}, S), \q B
= 1, 2, 3 , }
\end{equation}
where $\nu_{B}$, $\Theta$, $\varphi$, $\mu^{B}$ represent
subsidiary fields; $n$, $S$ and $\epsilon(n, S)$ denote the
particle's number, entropy and energy proper densities,
$J^{\alpha} = n u^{\alpha}$ is the particle current, and
$u^{\alpha}$ is the four-velocity, $u^{\alpha} = u^{0}(1,
\mathbf{v}/c)$, $u^{0} = 1/\sqrt{1 - \mathbf{v}^{2}/c^{2}}$; comma
denotes partial derivatives. Small Greek indexes run from $0$ to
$3$, and the Latin indexes run from $1$ to $3$; $x^{0} = ct$,
$\mathbf{r} = (x^{1}, x^{2}, x^{3})$. The metric tensor,
$g^{\alpha \beta}$, corresponds to the flat space-time in
Cartesian coordinates, $g^{\alpha \beta} = \text{diag} \{-1,
1,1,1\}$. The four-velocity obeys normalization condition
\begin{equation}\label{23_04_03}
u^\alpha u_\alpha = g_{\alpha \beta} u^\alpha u^\beta = -1 .
  \end{equation}
Below we consider the four-velocity and the particle density $n$
as dependent variables expressed in terms of the particles current
$J^{\alpha}$,
    \begin{equation}\label{23_02_04_1}
u^{\alpha} =  J^{\alpha}/|J|  , \q n = |J| =  \sqrt{-
J^{\alpha}J_{\alpha}} \, .
\end{equation}
The fluid energy obeys the second thermodynamic law
\begin{equation}\label{24_04_03}
  d \epsilon = nT d S + n^{-1} w d n \equiv nT d S + W d n \, ,
\end{equation}
where $T$ is the temperature and $w \eq \epsilon + p$ is the
proper enthalpy density, $p$ is the fluid pressure, $W = w/n$.

The action in \equ{26_04_3_0} depends (for a fixed or infinite
volume) on the independent variables $J^{\alpha}$,  $\Theta$, and
$Q = (\varphi, \mu^{B}, S)$, $A = A[J^{\alpha} , \varphi, \mu^{B}
\, , S, \nu_{B} \, , \Theta]$. Its variation results in the
following set of equations
   \begin{equation}\label{25_04_3_2}
\delta J^{\alpha} : \Longrightarrow W u_{\alpha} \eq V_{\alpha} =
- G Q_{,\alpha} \, ,
\end{equation}
    \begin{equation}\label{25_04_3_5}
\delta \varphi : \Longrightarrow J^{\alpha}_{\;\;, \alpha} = 0 ,
\end{equation}
    \begin{equation}\label{25_04_3_6}
\delta \mu^{B} : \Longrightarrow \partial_{\alpha} (J^{\alpha}
\nu_{B} ) = 0 , \q \text{or} \q D \nu_{A} =0,
\end{equation}
    \begin{equation}\label{25_04_3_7}
\delta \nu_{B}  : \Longrightarrow D \mu^{B} = 0 ,
\end{equation}
    \begin{equation}\label{25_04_3_3}
\delta S : \Longrightarrow  \partial_{\alpha} ( J^{\alpha} \Theta)
, \q \text{or} \q  D \Theta = - T ,
\end{equation}
    \begin{equation}\label{25_04_3_8}
\delta \Theta  : \Longrightarrow D S = 0 ,
\end{equation}
where $D \eq u^{\alpha} \p_{\alpha}$. \equ{25_04_3_2} gives us
Clebsch type velocity representation, cf. Ref. \cite{Brown93}.
Contracting it with $u^{\alpha}$ results in the dynamic equation
for the scalar potential $\varphi$,
\begin{equation}\label{28_03_04}
D \varphi = W .
\end{equation}
Both triplets $\mu^{B}$ and $\nu_{B}$ represent the advected
subsidiary fields and do not enter the internal energy. Therefore,
it is natural to treat one of them, say, $\mu^{B}$ as the flow
line labels.

Taking into account that the entropy and particles conservation
are incorporated into the set of variational equations, it is easy
to make sure that the equations of motion for the subsidiary
variables along with the velocity representation reproduces the
relativistic Euler equation. The latter corresponds to the
orthogonal to the flow lines projection of the fluid
stress-energy-momentum $T^{\alpha \beta}$ conservation, cf. Ref.
\ct{Mizner-Torn-Wheeler73,LL_73},
\begin{equation}\label{1}
 \frac{\partial T^{\alpha \beta}}{\partial x^{\beta}} \equiv T^{\alpha \beta}_{\;\;\;\;, \beta} = 0 ,
\end{equation}
\begin{equation}\label{en_mom}
T^{\alpha \beta} = wu^\alpha u^\beta + p g^{\alpha \beta} .
  \end{equation}
Note that the relativistic Euler equation can be written as
\begin{equation}\label{26_04_3_2}
( V_{\alpha , \beta} - V_{\beta, \alpha } ) u^{\beta} = T   S_{ ,
\alpha} \, ,
\end{equation}
where the thermodynamic relation
\begin{equation}\label{28_03_04_2}
d p = n d W - n T d S
\end{equation}
is taken into account. Vector $V_{\alpha}$, sometimes called Taub
current, \cite{Taub59}, plays an important role in relativistic
fluid dynamics, especially in the description of circulation and
vorticity. Note that $W$ can be interpreted as an injection energy
(or chemical potential), cf., for instance
\cite{Mizner-Torn-Wheeler73}, i.e., the energy per particle
required to inject a small amount of fluid into a fluid sample,
keeping the sample volume and the entropy per particle $S$
constant. Therefore, $V_{\alpha}$ is identified with the
four-momentum per particle of a small amount of fluid to be
injected in a larger sample of fluid without changing the total
fluid volume and the entropy per particle.

\paragraph{Boundary conditions.}
In order to complete the variational approach for the flows with
breaks, it is necessary to formulate the boundary conditions for
the subsidiary variables which do not imply any restrictions on
the physically possible breaks (the shocks, tangential and contact
breaks), are consistent  with the corresponding dynamic equations
and thus are equivalent to the conventional boundary conditions,
i.e., to continuity of the particle and energy-momentum fluxes
intersecting the break surface $R(x^{\alpha}) = 0$, cf. Ref.
\cite{LL_73},
\begin{equation}\label{29_03_04}
\{ \breve{J}\} = 0 , \q \breve{J} \eq J^{\alpha} n_{\alpha} \, ,
\end{equation}
\begin{equation}\label{29_03_04_1}
\{ T^{\alpha \beta} n_{\beta}\} = 0 ,
\end{equation}
where $n_{\alpha} $ denotes the unit normal vector to the break
surface,
\begin{equation}\label{29_03_04_2}
n_{\alpha} = N_{\alpha}/ N  , \q N_{\alpha} = R_{, \alpha} \, \q N
= \sqrt{N_{\alpha} N^{\alpha} } \, ,
\end{equation}
and braces denote jump, $\{ X\} \eq X|_{R = +0} - X|_{R = -0}$.

Our aim is to obtain boundary conditions as natural boundary
conditions for the variational principle. In the process of
deriving the volume equations we have applied integration by parts
to the term $J^{\alpha} G  \delta Q_{, \alpha}$. Vanishing of the
corresponding surface term along with that resulting from the
variation of the surface itself  lead to the appropriate boundary
conditions after the variational principle has been specified.

Rewriting the (volume) action with the break surface being taken
into account in the explicit form as
\begin{equation}\label{27_04_03_B6}
A = \int d^{4} x \sum_{\varsigma = \pm 1} \mathcal{L}^{\varsigma}
\theta (\varsigma R) \, ,
\end{equation}
where 
$\theta $ stands for the step-function, we obtain the residual
part of the (volume) action in the form
\begin{equation}\label{27_04_03_B6A}
\left. \delta A \right|_{res} = \int d^{4} x \sum_{\varsigma = \pm
1} \left[ \varsigma \mathcal{L} \delta_{D} ( R) \delta R +
\theta(\varsigma R) \p_{\alpha}  ( J^{\alpha} G  \delta Q )
\right] .
\end{equation}
Here $\delta_{D}$ denotes Dirac's delta-function and we omit index
$\varsigma$ labeling the quantities that correspond to the fluid
regions divided by the interface $R = 0$; superscript $\varsigma
\gtrless 0$ corresponds to the quantities in the regions $R
\gtrless 0$, respectively. Integrating the second term by parts
and supposing that the surface integral
    $
\int d^{4} x \sum_{\varsigma = \pm 1} \p_{\alpha}
\left(\theta(\varsigma R)  (u^{\alpha} G  \delta Q ) \right)
    $
vanishes due to vanishing of the variations $\delta Q$ at
infinity, we arrive at the residual action expressed by the
surface integral
\begin{equation}\label{30_04_3_10}
\left. \delta A \right|_{res} = \int d^{4} x \sum_{\varsigma = \pm
1} \varsigma \delta_{D} ( R) \left[ \mathcal{L} \delta R - R_{,
\alpha} J^{\alpha} G  \widetilde{\delta} Q \right] .
\end{equation}
$\widetilde{\delta} Q$ here means the limit values of the volume
variations, $\widetilde{\delta} Q^{\pm} \eq (\delta Q)_{R = \pm
0}$. It is convenient to express  these variations in terms of
variations of the boundary restrictions of the volume variables,
$\delta (X_{R = \pm 0}) \eq \delta\widetilde{X}^{\pm}$, and
variation of the break surface. It is easy to show that
\begin{equation}\label{1_05_3_7}
\widetilde{\delta} X = \delta \widetilde{X} + |N|^{-1} n^{ \alpha}
X_ {, \alpha} \delta R - X_ {, \alpha} P^{\alpha}_{\;\; \beta}
\delta f^{\beta} \,  ,
\end{equation}
where $P^{\alpha}_{\;\; \beta} = \delta^{\alpha}_{\;\; \beta} -
n^{\alpha}n_{ \beta}$, and $\delta f^{\beta}$ is an arbitrary
infinitesimal four-vector related to the one-to-one mapping of the
surfaces $R = 0$ and $R + \delta R = 0$.

Vanishing  of the action variation with respect to variations of
the surface variables $\delta R$ and $\delta f^{\beta}$ (which are
supposed to be independent) results in the following boundary
conditions
\begin{equation}\label{28_05_03}
\delta R : \Rightarrow   \left\{  p  + (u^{\alpha} n_{\alpha})^{2}
w \right\} = 0 ,
\end{equation}
\begin{equation}\label{28_05_03_1}
\delta f^{\beta} :  \Rightarrow P^{\gamma}_{\;\;\beta}  \left\{ W
J^{\alpha} N_{\alpha} u_{\gamma} \right\} = 0 , \;\; \mathrm{or}
\;\; P^{\gamma}_{\;\;\beta} \left\{ \check{J}  W  u_{\gamma}
\right\} = 0  ,
\end{equation}
which are equivalent to continuity of the momentum and energy
fluxes, cf. \equ{29_03_04_1}. Here we consider that the `on shell'
value of the volume Lagrangian density, $\mathcal{L}_{eq}$, is
equal to the pressure, $\mathcal{L}_{eq} = -\epsilon + n G D Q = -
\epsilon + w = p$.

Now we can complete formulation of the variational principle
appropriate both for continuous and discontinuous flows. The
independent volume variables are indicated above, and independent
variations of the surface variables are $\delta R$, $\delta
f^{\beta}$,  variations of the surface restrictions of the
generalized coordinates $\delta \varphi$, $\delta \mu^{B}$,
supposed to be equal from both sides of the break, $\{ \delta
\varphi \} = \{ \delta \mu^{B} \} =0$, and $\delta S$ with $\{
\delta S \} \ne 0$. Under these assumptions we arrive at the
following subset of the boundary conditions
\begin{equation}\label{28_05_03_2B}
\delta \widetilde{\varphi}: \Rightarrow    \{ J^{\alpha}
n_{\alpha}  \} \eq \{ \check{J} \}= 0  \quad \text{for} \quad \{
\delta \widetilde{\varphi} \} = 0 ,
\end{equation}
\begin{equation}\label{28_05_03_2C}
\delta \widetilde{\mu}^{B} : \Rightarrow   \{ \nu_{B} J^{\alpha}
n_{\alpha}  \} \eq \check{J} \{ \nu_{B} \} = 0  \quad \text{for}
\quad \{ \delta \widetilde{\mu}^{B} \} = 0 ,
\end{equation}
\begin{equation}\label{28_05_03_2}
\delta \widetilde{S}^{\pm} : \Rightarrow J^{\alpha}
n_{\alpha}\widetilde{ \Theta}^{\pm} \eq \check{J} \widetilde{
\Theta}^{\pm} = 0 .
\end{equation}
Eqs.~\eqref{28_05_03}--\eqref{28_05_03_2B} reproduce the usual
boundary conditions, and Eqs.~\eqref{28_05_03_2C},
\eqref{28_05_03_2} are the boundary conditions for the subsidiary
variables. Other boundary conditions for the latter variables do
not strictly follow from the variational principle under
discussion. But we can find them from the corresponding volume
equations of motion, providing, for instance, that they are as
continuous as possible.\footnote{Note that choice of the boundary
conditions for the fields $\varphi$, $\mu^{B}$, $\nu_{B}$ and
$\Theta$ is not unique due to the fact that they play roles of the
generalized potentials and therefore possess the corresponding
gauge freedom relating to the transformations $\varphi, \mu^{B},
\nu_{B}, \Theta \rightarrow \varphi', \mu'^{B}, \nu'_{B}, \Theta'$
such that $u'_{\alpha} = u_{\alpha}$ (given by the representation
\eqref{25_04_3_2}). For instance, it seems possible to use entropy
$S$ as one of the flow line markers. But if we are dealing with
discontinuous flows then it is necessary to distinguish the
Lagrange markers of the fluid lines, $\mu^{B}$, and the entropy,
$S$. Namely, the label of the  particle intersecting a shock
surface evidently does not change, but the entropy does change.
Thus, entropy can be chosen as one of the flow line markers only
for the flows without entropy breaks.} The natural choice
corresponds to continuity of their fluxes,
\begin{equation}\label{27_04_03_C13}
  \{  n_{\alpha} u^{\alpha}  n \mu^{B}  \} \eq \check{J} \{ \mu^{B}  \} =0  ,
\end{equation}
\begin{equation}\label{27_04_03_C14}
   \{ n_{\alpha} u^{\alpha}  n \varphi  \} \eq \check{J} \{ \varphi  \} = 0 .
\end{equation}
The set of the boundary conditions given by
Eqs.~\eqref{28_05_03}--\eqref{27_04_03_C14} is complete and allows
one to describe any type of breaks, including shocks. For the
latter case $\check{J} \ne 0$ and we arrive at continuity of the
variables $\nu_{B}$, $\mu^{B}$, $\varphi$ and zero boundary value
of $\Theta$. For $\check{J} = 0$ the flow lines do not intersect
the break surface and we obtain very weak restrictions on the
boundary values of the subsidiary variables, cf. nonrelativistic
case discussed in Refs.~\cite{KK_97,Kats_01}. Note that for the
specific case $\check{J} = 0$ (slide and contact discontinuities)
we can simplify the variational principle assuming all both-side
variations of the subsidiary variables to be independent.

The above variational principle allows modifications. First, it is
possible to exclude constraints, expressing the four-velocity by
means of representation \eqref{25_04_3_2}. In this case the volume
Lagrangian density can be chosen coinciding with the fluid
pressure, cf. Ref. \cite{Brown93}, where the continuous flows are
discussed in detail. Second, we can include into the action the
surface term respective for the surface constraints, cf.
Refs.~\ct{KK_97,Kats_01,KATS_02,KATS_02A}, where such surface
terms are discussed for ideal hydrodynamics and
magnetohydrodynamics in the nonrelativistic limit. This can be
done for the cases both with excluded and non excluded volume
constraints.

\textsc{\textbf{Canonical variables.}} Starting with the action of
\equ{26_04_3_0} and Lagrangian density given by \equ{23_02_04_4}
we can introduce the canonical (Hamiltonian) variables according
to the general receipt. Namely, let $Q$ represents the canonical
coordinates. Then
\begin{equation}\label{2_02_04}
P \eq \frac{\delta A}{\delta Q_{,0}} = J^{0} G \eq (\pi_{\varphi}
\, , \pi_{\mu^{B}}, \pi_{S})
\end{equation}
gives us conjugate momenta. Relations \eqref{2_02_04} cannot be
solved for the generalized velocities $Q_{,0}$ suggesting that we
are dealing with the degenerated (constraint) system, cf. Refs.
\cite{Dirac64,Gitman86,GP93,ZakharovKuznetsov97}. But the
constraints are of the first type. Thus,  performing the Legendre
transform with respect to $Q$ we arrive at the Hamiltonian density
\begin{equation}\label{2_02_04_2}
\mathcal{H} =  P Q_{,0} - p(W,S)  ,
\end{equation}
where we suppose that the four-velocity is given by representation
\eqref{25_04_3_2}. Making use of the definition \eqref{2_02_04}
and of the time component of the velocity representation,
\equ{25_04_3_2}, we can transform the first term in
\equ{2_02_04_2} as
\begin{equation}\label{2_02_04_2B}
P Q_{,0} = J^{0} G Q_{,0} = - \pi_{\varphi} V_{0} = \pi_{\varphi}
V^{0} .
\end{equation}
Taking into account the normalization condition for the Taub
current, $V_{\alpha}V^{\alpha} = - W^{2}$, we obtain
\begin{equation}\label{2_02_04_2C}
V^{0} = \sqrt{W^{2} + V_{a}V^{a}} \, .
\end{equation}
Consequently, we arrive at the following Hamiltonian density
\begin{equation}\label{2_02_04_2A}
\mathcal{H} \eq \mathcal{H}(P,Q, Q_{, a}; W) = \sqrt{W^{2} +
V_{a}V^{a}} \, \pi_{\varphi} - p(W, S) .
\end{equation}
In terms of the canonical coordinates and momenta the space
components of the velocity are
\begin{equation}\label{24_05_04}
\pi_{\varphi} V_{a} = - P Q_{,a} .
\end{equation}
The canonical equations following from this Hamiltonian reproduce
in a $3+1$ form the above dynamic equations for the variables
entering the Taub current representation. Variation of the action
with respect to the chemical potential $W$ results in
\begin{equation}\label{22_02_04_4}
n = \frac{\pi_{\varphi}}{\sqrt{1 + V_{a}V^{a}/W^{2}}}   \, .
\end{equation}
Obviously, this relation is equivalent to \equ{2_02_04_2C},
expressing the particle density $n$ in terms of the variables
entering the Hamiltonian.

Underline, that the Hamiltonian given by \equ{2_02_04_2A} depends
not only on the generalized coordinates $\varphi$, $\mu^{B}$, $S$,
their spatial derivatives and conjugate momenta, but also on the
chemical potential $W$ as well.   Evidently, we can consider $W$
as the additional generalized coordinate with zero conjugate
momentum, $\pi_{W} = 0$. This condition is consistent with the
dynamic equations due to the fact that $\p_{0} \pi_{W} = \p
\mathcal{H}/ \p W = 0$, cf. \equ{22_02_04_4}.

Bearing in mind the flows with breaks one can see that in the
above discussed variant of the least action principle we do not
arrive at the additional surface variables except that defining a
break surface, $R$. But it enters the action functional without
derivatives. Therefore, corresponding conjugate momentum is
zero-valued. Introducing the Hamiltonian variables for the flows
with breaks we have to treat $R$ as the surface function, defining
some (surface) constraint. The latter is nothing else than
continuity of the normal component of the fluid momentum flux,
\equ{28_05_03}.

\paragraph{Poisson brackets.}\label{Poisson brackets}

The Poisson brackets in the canonical variables are of a standard
form. Symbolically,
\begin{equation}\label{Poisson br1}
\{Q^{\mathcal{A}}(x) , P_{\mathcal{B}}(y) \} =
\delta^{\mathcal{A}}_{\mathcal{B}} \delta(x - y),
\end{equation}
where $\delta(x - y)$ is spacetime Dirac's delta, $x $ and $y$ are
spacetime points, $Q^{\mathcal{A}}$ is shorthand notation for
$\varphi$, $\mu^{A}$ and $S$, and, analogously,  $P_{\mathcal{B}}$
denotes corresponding conjugate momenta, $\pi_{\varphi}$,
$\pi_{\mu^{B}}$ and $\pi_{S}$.

\paragraph{Ertel invariant.}
In addition to energy, momentum, and angular momentum
conservation, for the ideal hydrodynamic flows there are exist
specific local conservation laws related to the dragged and
frozen-in fields, and corresponding  topological invariants
(vorticity, helicity, Ertel invariant, etc.), cf. Refs.
\cite{lamb,Salmon82,STY_90,GP93,ZakharovKuznetsov97} and citations
therein for the nonrelativistic case. They are caused by the
relabeling symmetry. Discussion of these problems for the
relativistic flows seems insufficient, see Refs.
\cite{Taub59,Schutz70,Brown93,ZakharovKuznetsov97} and citations
therein. Exploitation of the above description permits one
considering these invariants to be simplified. For example,
consider here generalization of the Ertel invariant for the
relativistic fluids (to my best knowledge, this item was not
discussed earlier). Defining the Ertel four-current,
\begin{equation}\label{9_11_03}
\mathcal{E}^{\alpha} = - \frac{1}{2} \epsilon^{\alpha  \beta \mu
\nu } \omega_{ \beta  \mu } S_{ ,\nu } =  - {^{\ast}
\hspace{-0,2em}\omega}^{\alpha  \nu } S_{ ,\nu }  \, ,
\end{equation}
one can see that it is divergence-free,
$\mathcal{E}^{\alpha}_{\;\; ,\alpha} = 0$. Here $\epsilon^{\alpha
\beta \mu \nu }$ is Levi-Civita tensor, $\omega_{ \beta \mu }$ is
the (Khalatnikov) vorticity tensor, $\omega_{ \beta  \mu } =
V_{\mu , \beta} - V_{\beta , \mu}$, and ${^{\ast}
\hspace{-0,2em}\omega}^{\alpha \nu }$ is its dual. Moreover, the
Ertel four-vector $\mathcal{E}^{\alpha}$ is proportional to the
particle current $J^{\alpha}$,
\begin{equation}\label{15_02_04}
\mathcal{E}^{\alpha} = \mathbb{E} J^{\alpha} ,
\end{equation}
in view of $\mathcal{E}^{\alpha}_{\;\; ,\alpha} = 0$ resulting in
$\mathbb{E} \eq \mathcal{E}^{0}/J^{0}$ being dragged by the fluid,
\begin{equation}\label{15_02_04_1}
D \mathbb{E} = 0  ,
\end{equation}
i.e. $\mathbb{E}$ is the scalar invariant of the motion. In the
nonrelativistic limit it coincides with the Ertel invariant $(\cu
\mathbf{v} \c \bn s)/\rho$, where $\rho$ denotes the fluid
density.

\paragraph{Helicity current.}
The helicity invariant in the nonrelativistic case exists for the
barotropic flows and presents pseudoscalar $ \mathbf{v} \c \cu
\mathbf{v}$. The strict analog for the relativistic case is the
pseudovector
\begin{equation}\label{9_11_03_1}
Z^{\alpha} = \frac{1}{2}\epsilon^{\alpha \beta \mu \nu } \omega_{
\beta \mu } V_{ \nu } \eq {^{\ast} \hspace{-0,2em}\omega}^{\alpha
\nu } V_{ \nu } \, ,
\end{equation}
Strict calculations show that for the isentropic flows the
helicity current $Z^{\alpha}$ is conserved, $Z^{\alpha}_{\;\;
,\alpha} = 0$.

For the general type flows there exists generalization of the
helicity current. Namely, consider reduced Taub vector,
\begin{equation}\label{26_01_04}
\widetilde{V}_{\alpha}  \eq V_{\alpha} + \Theta S_{, \alpha}  \, ,
\end{equation}
where $\Theta $ obeys \equ{25_04_3_3}, and the corresponding
reduced vorticity tensor,
\begin{equation}\label{26_01_04_1}
\widetilde{\omega}_{ \alpha \beta} \eq \widetilde{V}_{\beta ,
\alpha} - \widetilde{V}_{\alpha , \beta} \, .
\end{equation}
This tensor is orthogonal to the flow lines,
\begin{equation}\label{26_01_04_5}
\widetilde{\omega}_{ \alpha \beta} u^{\beta} = 0,
\end{equation}
and the reduced helicity current
\begin{equation}\label{26_01_04_7}
\widetilde{Z}^{\alpha}  = {^{\ast}
\hspace{-0,1em}\widetilde{\omega}}^{\alpha \nu } \widetilde{V}_{
\nu }
\end{equation}
is conserved for arbitrary flows,
\begin{equation}\label{26_01_04_8}
\widetilde{Z}^{\alpha}_{\;\; , \alpha}  = {^{\ast}
\hspace{-0,1em}\widetilde{\omega}}^{\alpha \nu }  \widetilde{V}_{
\nu , \alpha}  = \frac{1}{2} {^{\ast}
\hspace{-0,1em}\widetilde{\omega}}^{\alpha \nu }
\widetilde{\omega}_{\alpha \nu  } = 0  .
\end{equation}

\paragraph{Conclusion.}

We have shown that it is possible to describe the relativistic
ideal fluids with all physically allowable breaks in terms of the
least action principle both in the Lagrangian and Hamiltonian
description. The boundary conditions for the subsidiary variables
entering the Clebsch type velocity representation are obtained in
two different ways: one part follows from the variational
principle as natural boundary conditions while the other one was
obtained from the dynamic equations under assumption relating to
absence of the corresponding sources and the maximal continuity
compatible with the volume equations. Note that it is possible to
change the variational principle in such a way that all boundary
conditions will result from it, i.e., they become natural boundary
conditions. For this purpose it is necessary to modify the
variational principle by adding a surface term with corresponding
constraints, similarly to the nonrelativistic case (compare with
the papers \cite{KK_97,Kats_01} for the hydrodynamics and
\cite{KATS_02,KATS_02A} for the magnetohydrodynamics). This
variants are to be discussed in the forthcoming papers.

The approach discussed allowed us to give a simple treatment of
the additional invariants of the motion, in particular, to present
generalization of the Ertel invariant for the relativistic flows.
This approach is suitable for the general relativity and for the
relativistic magnetohydrodynamics as well. Note that for the flows
without breaks the general relativity case is discussed in detail
in the paper \cite{Brown93}. The discontinuous flows for the
general relativity can be described in analogy to the above
discussion and will be published elsewhere.

\subsection*{Acknowledgment}

\frenchspacing

This work was supported by the INTAS (Grant No. 00-00292).

\end{document}